# Temporal contrast reduction techniques for high dynamic-range single-shot temporal contrast measurement

Xiong Shen[1], Peng Wang[1], Jingxin Zhu[1,2], Zhe Si[1,2,3], Yuxia Zhao[4], Jun Liu[1,2,5,6*] and Ruxin Li[1,2,5,6*]

[1]State Key Laboratory of High Field Laser Physics, Shanghai Institute of Optics and Fine Mechanics, Chinese Academy of Sciences, Shanghai 201800, China

[2]University Center of Materials Science and Optoelectronics Engineering, University of Chinese Academy of Sciences, Beijing 100049, China

[3]Department of physics, shanghai University, Shanghai 200444, China

[4]Technical Institute of Physics and Chemistry, Chinese Academy of Sciences, Beijing 100190, China

[5]CAS Center for Excellence in Ultra-intense Laser Science (CEULS), Shanghai Institute of Optics and Fine Mechanics, Chinese Academy of Sciences, Shanghai 201800, China

[6]IFSA Collaborative Innovation Center, Shanghai Jiao Tong University, Shanghai 200237, China

*Corresponding author E-mail: jliu@siom.ac.cn, ruxinli@siom.ac.cn

**Abstract**

A single-shot characterization of the temporal contrast of a petawatt laser pulse with a high dynamic-range, is important not only for improving conditions of the petawatt laser facility itself, but also for various high-intensity laser physics experiments, which is still a difficult problem. In this study, a new idea for improving the dynamic-range of a single-shot temporal contrast measurement using novel temporal contrast reduction techniques is proposed. The proof-of-principle experiments applying single stage of pulse stretching, anti-saturated absorption, or optical Kerr effect successfully reduce the temporal contrast by approximately one order of magnitude. Combining with the SRSI-ETE method, its dynamic-range characterization capability is improved by approximately one order of magnitude to approximately $10^9$. It is expected that a higher dynamic-range temporal contrast can be characterized using cascaded temporal contrast reduction processes.

**1. Background**

The development of chirped pulse amplification (CPA)[1] and optical parametric chirped pulse amplification (OPCPA)[2] techniques has increased the peak laser power to petawatt (PW) level since the first demonstration of laser[3]. Many PW laser systems have been demonstrated in national or even university laboratories worldwide.[4-8] Several 10 PW-level laser facilities, such as APOLLON 10-PW,[9] ELI-NP,[10] Vulcan-10 PW,[11] SULF-10 PW,[12] and PEARL-10 PW[13] are being constructed. Even 100 PW-level laser facilities, such as SEL,[14] OPAL,[15] XCELS,[16] and ELI[17] are proposed to be constructed in the near future, where the SEL facility has been launched in 2018.[14] The focal intensity of these PW laser pulses will reach $10^{20}$-$10^{23}$ W/cm$^2$, which is much higher than the so-called relativistic optics regime of $10^{18}$ W/cm$^2$. Such high intensity lasers provide previously unavailable extreme conditions in laboratories for many specific important research activities,[18] including the generation and acceleration of electrons,[19-21] protons,[22] and ion particles,[23] laboratory astronomy,[24] fast-ignition inertial confinement fusion,[25] the generation of a secondary source of high-intensity γ-rays[26] or even vacuum birefringence,[27] etc., which in recent years have been a hot topic of research – the physics of ultrahigh-intensity lasers.

Many previous studies have proved that the results of laser-plasma interaction are not only related to the focal intensity, but also to the temporal contrast of the laser pulse.[28-29]. The expected results of laser-plasma interaction can be even worse when an ultra-high intensity laser with a lower temporal contrast is compared with a relatively low intensity laser pulse with a higher temporal contrast. This is because the pre-pulses in such an ultrahigh-intensity laser would be high enough to ionize the target to generate pre-plasmas that could destroy the experiment. Therefore, high temporal contrast is vital for the application of such ultra-intense laser pulses.

Until now, the temporal contrast of PW laser pulses could be significantly improved using several techniques, such as the OPCPA front amplifier stage,[30] the double CPA setup based on intermediate pulse cleaning using saturated absorption,[31] optical parametric amplification,[32-33] polarization rotation,[34] self-diffraction (SD),[35-36] cross-polarized wave (XPW) generation,[37-38] etc. These improve the temporal contrast of seed pulses up to $10^{11}$ or higher. An in-band noise reduction method after the amplifier has recently been proposed to improve temporal contrast.[39] A plasma mirror can also be used after the compressor to improve the temporal contrast.[40]

To study the improvement in temporal contrast of an ultra-intense pulse or the relationship of temporal contrast with experimental results, an important first step is to precisely characterize the temporal contrast of a pulse. As for the PW laser systems, the laser repetition rate is low or even single shot; therefore, the characterization of the temporal contrast of PW laser pulses has to work in the single-shot mode. Although great progress has been made in recent years in enhancing the temporal contrast of ultra-intense laser pulses, there are fewer techniques for the single-shot characterization of the temporal contrast of a pulse with both a high dynamic-range and a high temporal resolution.

The first single-shot measurement of temporal contrast using the cross-correlator technique was reported in 2001,[41] and many improvements were made after this.[42-45] The main idea of the cross-correlator is to encode time into space, and then the temporal contrast can be obtained from the spatial intensity distribution on the detectors. In this technique, the temporal resolution is limited by the resolution of the detectors, the group velocity mismatch, which is determined by the cross angle of the two incident pulses, and the thickness of the nonlinear crystal, while the dynamic-range is limited by the signal energy, sensitivity and dynamic-range of the detectors. The highest dynamic-range known for a single-shot temporal contrast measurement is $10^{10}$ which is achieved by reducing the signal from the main peak pulse using a dot or strip-shaped density filter, where the temporal resolution is approximately 700 fs or 160 fs, respectively.[44-45] In principle, it is possible to increase the signal energy by increasing the energy of the input pulse; however, this is limited by the thickness, damage threshold, and size of the nonlinear crystal.

A technique based on the idea of encoding time to frequency can also be used for single-shot temporal contrast characterization with a temporal resolution as precise as sub-20 fs.[46] Due to the property of the heterodyne method, a single-shot temporal contrast measurement of a $10^5$ dynamic range using self-referenced spectral interferometry (SRSI) was first reported in reference.[47] A single-shot $10^6$ dynamic-range was measured using the modified SRSI technique in 2013.[48] Recently, an improved method, the SRSI-extended time excursion (SRSI-ETE) method, has been found to increase the width of the time window by almost two-fold to approximately 18 ps.[46] However, due to the limitation of the signal-to-noise ratio and the dynamic-range of the detector, the dynamic-range is limited to $10^8$.

The dynamic-range and the signal-to-noise ratio of detectors such as CCD, sCMOS, PMT, etc., are key parameters for single-shot temporal contrast measurements reaching a high dynamic-range. Currently the highest dynamic-range of all available detectors is approximately $10^4$, which cannot be improved. For a cross-correlator, the dynamic-range of a temporal contrast single-shot measurement can be improved by increasing the pulse energy of the signal and then attenuating the signal of the main peak pulse.[44-45] This improvement allows the measurement to reach a maximum dynamic-range of about $10^{10}$ so far, which is not high enough for laser systems of tens to hundreds of PW. As for the SRSI-ETE method, the signal generated by the main peak pulse cannot be attenuated. Therefore, developing a method to further improve the dynamic-range of both techniques for single-shot temporal contrast measurement, especially for SRSI-ETE, is an important goal.

In this study, we propose a new idea to improve the dynamic-range of the temporal contrast characterization of a laser: temporal contrast reduction (TCR) of the pulse firstly, and then measure the decreased temporal contrast by using the above existing methods. As proof-of-principle experiments, the pulse stretching, anti-saturation absorption, and the optical Kerr effect (OKE) are employed to help decrease the temporal contrast of a laser pulse. The results show that these three methods can work well and that the temporal contrast can be decreased by approximately one order of magnitude with the application of only one stage of the process. We expect that a decrease of several orders of magnitude can be achieved using cascaded processes or a combination of several processes in the future. Therefore, we expect that the dynamic-range of a single-shot temporal contrast measurement could be improved to as much as $10^{11-12}$ by applying the proposed TCR method and the above-mentioned cross-correlator or SRSI-ETE techniques together.

**2. Principle**

The principle of the proposed idea is shown in **figure 1**. In the first step, a pre- or post-pulse is introduced before or after the main peak pulse as a calibrating pulse for TCR. This can be easily implemented by placing a glass plate with a fixed thickness in the optical path. In principle, the temporal contrast of the test pulse can be decreased in two ways: 1) weakening the main peak pulse (WMPP) while keeping the rest unchanged, and 2) amplifying the pre-pulses and/or background noise (AMPB) from ASE or parametric fluorescence while keeping the main peak pulse unchanged. Both methods can also be combined to further decrease the temporal contrast. After TCR, the temporal contrast of the modulated test pulse is characterized by either the cross-correlator or SRSI-ETE techniques discussed previously. Finally, the high dynamic temporal contrast measurement is reconstructed by combining the TCR value and the measurement result of the modulated test pulse.

Then, how to realize WMPP or AMPB to decrease the temporal contrast of the test pulse? As for the WMPP, it is natural to assume that this should be a fast process due to a nonlinear dependence on the pulse intensity. Of course, the second-harmonic generation (SHG) is such a process. However, the main peak pulse can only be weakened by less than 50% due to the limited energy transfer efficiency in a single stage of SHG. Cascaded SHG processes can further decrease the temporal contrast by one order of magnitude; however this complicates the optical setup. Inspired by the techniques used in optical limiting researches,[49-50] anti-saturation absorption and OKE are used in the proof-of-principle experiments for this study. As for AMPB, optical parametric amplification of the pre-pulses and background noise would be a common approach. However, this requires relatively high energy and long pulse duration for the pump beam. Therefore, we will not discuss any AMPB methods here. Except for WMPP and AMPB, we will introduce a very simple method — pulse stretching, which can decrease both the main peak pulse, and pre- or post- pulses, while maintaining background noise.

## 3. Experimental results and discussion

### 3.1. Pulse stretching

In the temporal domain, an ultrashort pulse can be simply classified into three parts: the main peak pulse, pre- or post- pulses, and background noise (ASE or parametric fluorescence). Pre- or post-pulses have almost the same level of pulse duration and spectrum as those of the main peak pulse, because they originate from the front-and-back reflection of parallel optical plates. In general, the temporal intensity ratio between the main peak pulse $I_{main}$ and the background noise $I_{BG}$ is defined by the dynamic-range of the temporal contrast of an ultrashort pulse, which is the most difficult parameter to characterize.

For a Gaussian laser pulse, the temporal intensity of a pulse is closely related to its full width at half maximum (FWHM) pulse duration, which can be simply and approximately described as $I = E/\tau$, where $E$ is the pulse energy and $\tau$ is the FWHM pulse duration. For a pulse with certain energy $E$, if the pulse is stretched or chirped in the temporal domain, the temporal intensities $I$ of the main peak pulse and the pre- or post-pulses will linearly decrease with increasing pulse duration $\tau$ while that of the background noise remains the same. Here, the intensity ratio between the main peak pulse and the pre- or post-pulses will not change, but the intensity ratio between the main peak pulse and the background noise will decrease. In this way we can decrease the temporal contrast by simply chirping the laser pulse to a relatively long duration.

We performed a proof-of-principle experiment using a Ti: sapphire CPA laser system running at 800 nm central wavelength and 1 kHz repetition rate, which can stretch the pulse duration through a grating-based pulse compressor. Laser pulses of about 3 mJ running at three different pulse durations were characterized separately using a commercial delay-scanning third-order cross-correlator (Sequoia 800, Amplitude) with a 17 fs scan step. **Figure 2** (a) shows the third-order autocorrelation curves of the three pulses. The FWHM of the three curves are approximately 240 fs (pulse 1), 670 fs (pulse 2), and 1476 fs (pulse 3). The pulse duration ratio of these three pulses is calculated to be $r_1$ = 1: 2.79: 6.15. The temporal contrast curves of the three pulses measured by Sequoia 800 are shown in figure 2 (b). It can be seen that the intensities of pre- or post-pulses are almost the same at different temporal resolutions for all three pulses, whereas the intensity of the background noise increases with stretched pulse duration. For example, at -6 ps (position *a* marked in figure 2 (b)), the intensities are $9.7 \times 10^{-8}$, $2.7 \times 10^{-7}$, and $5.9 \times 10^{-7}$ according to pulse 1, 2, and 3, respectively; the intensity ratio of the background noise at these three pulse widths is $r_2$ = 1: 2.78: 6.08, which agrees very well with $r_1$. It can then be concluded that there is a linear relationship between the pulse stretching and the TCR of the background noise.

The experimental results confirm that the TCR method using pulse stretching is very simple; we only need to stretch the femtosecond pulse by tuning the distance of a real or mirrored parallel grating in the compressor. The pulse duration can be monitored using any pulse duration measurement instrument (such as an autocorrelator) or obtained from the correlation curves of the temporal contrast measurement device itself. For the case where the PW laser pulse duration is less than 30 fs, pulse stretching of at least two orders of magnitude can be applied, stretching the pulse up to approximately 3 ps, provided that the pulse energy is sufficient for pulse characterization. As a result, the temporal contrast between the main peak pulse and the background noise is decreased by two orders of magnitude. Of course, the precise structures and positions of pre- or post-pulses cannot be refined because of the decreased temporal resolution, as shown in figure 2 (b). However, because pre- or post-pulses are stronger by several orders of magnitude than background noise in general, the temporal contrast of these pulses can be measured using an existing method and instrument with a relatively low dynamic-range. Therefore, high dynamic temporal contrast can be measured in two steps: 1) obtain the temporal contrast of background noise using a stretched pulse, and 2) achieve the precise structures of pre- or post-pulses using conventional methods with high temporal resolution but relatively low dynamic-range. Using this simple pulse stretching method, a commercial delay-scanning third-order cross-correlator can also improve its capability in the dynamic-range by one or two orders of magnitude.

### 3.2. Anti-saturation absorption and optical Kerr effect

Although pulse stretching is a simple and practical method for TCR, temporal resolution is sacrificed where the pulse energy of the main peak pulse does not decrease directly. However, the pulse-stretching method can be used to find the ratio between the main peak pulse and the background noise of a pulse with high temporal contrast; the precise structures and positions of pre- or post-pulses would need to be characterized by other conventional methods for the second step above. Therefore, direct reduction of the pulse energy of the main peak pulse is desired to simplify the temporal contrast measurement to one step.

Is there any process that can reduce the pulse energy of the main peak pulse? We found that a lot of research exists in the field of optical limiting with the same purpose of weakening a strong laser for laser protection. This approach can be applied to the TCR method; the process must be ultrafast with the shortest picosecond decay time possible so that the main peak pulse is well separated from the background noise and

pre- or post-pulses. We found that the two-photon absorption based anti-saturation absorption process and OKE satisfy this requirement very well.[49-50] In the following proof-of-principle experiments, we will show the application of both methods to TCR.

A Ti: sapphire CPA laser system (Legend Elite Series, Coherent) with energy of approximately 4 mJ/40 fs /1 kHz/ 800 nm was used in both of the proof-of-principle experiments. Due to the limited pulse energy of the kHz Ti:sapphire laser system, the temporal contrasts of the laser pulse were characterized using the novel SRSI-ETE method[46] for both the original pulse and the TCR pulse. The temporal contrast of the original pulse was also verified using the commercial delay-scanning third-order cross-correlator (Sequoia 800, Amplitude). It should be noted that SRSI-ETE is also a pulse duration characterization method which is based on spectral interferometry; it has a very high sensitivity to the pulse energy due to its linear property.[51] Therefore, SRSI-ETE is an ideal technique for the temporal contrast characterization of weak pulses. The general diagram of a TCR-based SRSI-ETE device is shown in **figure 3** (a), where the reference pulses are generated using either XPW generation,[52] transient grating (TG),[53] or SD process.[54] Here, the SD process was used to generate a reference pulse for SRSI-ETE measurement.

The diagram and optical setup of the experiments using the SD process is shown in figure 3 (b). The input pulse to be characterized was split into two beams, the reflective arm was used for generating a reference pulse for the SRSI-ETE through the nonlinear SD process, and the transmitted pulse was used as a test pulse. The nonlinear SD process was used for the generation of the reference pulse because it is a spatially well-separated four-wave mixing process. The temporal contrast of the first-order SD signal was almost a cube of contrast for the incident pulse.[36] The reflective beam was initially adjusted through a variable neutral density (VND) filter. The beam was then split into two beams, a time delay stage Delay 1 was used to adjust the time delay between the two beams, and two lenses with the same focal length were used to focus the two beams on the Kerr medium K. A fused silica wedge plate with a thickness of approximately 0.2 mm and a wedge angle of 2 degrees was used as the Kerr medium; note that SD signals are generated when two incident beams overlap the Kerr medium in time and space. There was no parallel transparent optical element on the optical path of the reference beam which prevented post-pulses introducing from reflection through the front and rear surfaces.

The test pulse first passed through the transparent plate *P* to introduce a replica of the main peak pulse as a calibrating pulse. Then a TCR process was carried out to decrease the temporal contrast of the test pulse. Next the collimated reference pulse and the TCR test pulse were focused in the entrance slit of the 2D imaging spectrometer with a small crossing angle between them in the vertical direction using a reflective cylindrical mirror C2. The time delay between these two pulses was adjusted by the delay stage Delay 2, and the pulse energy of the reference pulse and the TCR test pulse were adjusted by two VND filters, VND 1 and VND 2, respectively. With an appropriate time delay, a tilted spectral interferogram was recorded by the 2D imaging spectrometer (SP2750, Princeton Instruments) and a 2048 × 512 pixel CCD camera (PIXIS 512, Princeton Instruments). The Fourier transform of the interferogram in both the spectral domain and spatial domain will obtain separated DC and AC terms of the interferogram in the spatial frequency domain. Then, the temporal contrast of the TCR test pulse was calculated by filtering the AC term.

We assumed a plate P with a thickness *L* and reflectivity *R1* and *R2* for the front and rear surface, respectively. A post-pulse for the delay time *t = 2nL/c* away from the main peak pulse was then introduced for the perpendicular arrangement of the plate P, where *n* is the refractive index and *c* is the velocity of light in vacuum. The intensity of the calibrating post-pulse relative to the main peak pulse is *I = R1\*R2*. The normalized value of the increased intensity of the calibrating pulse after TCR compared to that of the original test pulse (known from the traditional temporal contrast characterization) determines the reduction amount of the temporal contrast for the TCR pulse.

### 3.2.1. Anti-saturation absorption effect

It is well known that the saturated absorption effect which transmits the main peak pulse while stopping the weak pre- or post-pulses has long been used to improve the temporal contrast.[55] The temporal contrast was improved by more than two orders of magnitude in a PW laser system using a saturated absorption glass plate.[31] Contrary to the saturated absorption effect, the anti-saturated absorption process absorbs the strong main peak pulse and transmits the weak pre- or post-pulses and background noise. Therefore, anti-saturated absorption can be used to decrease the temporal contrast.

Few materials are suitable for the optical limitation of femtosecond laser pulses. It has been experimentally proven that an anti-saturation absorption material named 2-[Bis-(4'-(di(2,5,8,11,14-pentaoxahexadecan-16-yl)amino)-bipheny-4-yl)-methylene]-malononitrile (LBDBP) can be used to absorb a high-intensity femtosecond pulse.[56] LBDBP has been successfully used to stabilize the pulse energy of a high-intensity laser pulse because its fast response time.[56] Therefore, LBDBP was used in our proof-of-principle experiment setup as shown in **figure 4**. Here a laser pulse with energy of approximately 3.5 mJ at 800 nm was used, and the VND filter adjusted the pulse energy of the input beam. A beam splitter (BS) with an R/T ratio of 7:93 split the input beam into two beams. After the beam splitter, the reflected beam was used for reference pulse generation through the SD process; the transmitted beam was passed through a 0.5 mm thick fused silica plate P to obtain a calibrating pulse and then focused by a lens with a focal length of 500 mm. The beam diameter was reduced from the original 12 mm to about 5 mm on the LBDBP sample, which is located approximately 200 mm in front of the focal point. Pure LBDBP material (solid at a temperature of 23 ℃) was poured into a fused silica cuvette with a 1 mm path length and 1 mm wall thickness. The cuvette with LBDBP was then heated to 85 ℃ and maintained at this temperature throughout the experiment.

The optical limiting property of LBDBP was measured at the beginning of the experiment. The input power on to the LBDBP sample was adjusted using the VND filter; the output power from the LBDBP sample was measured by a power meter. **Figure 5** (a) shows the relationship between the transmissivity and the input laser intensity on the sample. It was found that the output power, or transmissivity, decreased rapidly as the laser intensity on the sample increased from zero to approximately 5 GW/cm². Then, the transmissivity slowly decreased from approximately 19% to approximately 17% when the laser intensity increased from 5 GW/cm² to approximately 19 GW/cm². This is exactly the property of the anti-saturation absorption effect. Usually, the intensity of pre- or post-pulses and background noise is typically several orders of magnitude lower than that of the main peak pulse. As a result, the main peak pulse is reduced by approximately five times more than the pre- or post-pulses, and the temporal contrast decreases.

At the beginning, the temporal contrast of the input test pulse was characterized without the LBDBP sample using both the commercial third-order cross-correlator (Sequoia 800) and our device based on the SRSI-ETE method as depicted in figure 4. The results are shown in figure 5 (b). Both measurements show good correlation with each other, which confirms the reliability of our SRSI-ETE device for the temporal contrast measurement.

The LBDBP sample was then added into the fused silica cuvette in the optical path. The temporal contrast of the TCR pulse using the anti-saturated absorption effect was characterized using our device based on the SRSI-ETE method; the laser intensity on the surface of LBDBP was approximately 19 GW/cm². The temporal contrast of the TCR pulse through the LBDBP sample is shown in figure 5 (b) as the red solid curve. It should be noted that calibrating pulses located at $t$ = 5 ps were introduced by a 0.5 mm thick fused silica plate P. The normalized pulse intensity of the calibrating pulse changed from 1.97 × 10$^{-4}$ to 8.65 × 10$^{-4}$ with 4.4-fold increase. As shown in figure 5 (a), a transmissivity of 16.7% at an intensity of 19 GW/cm$^2$ represents approximately six times the TCR due to the optical limiting effect; it is assumed that the pre- or post-pulse and background noise are transmitted at 100%. Moreover, the surface reflection of the fused silica cuvette and the linear absorption of the LBDBP sample made the transmission of pre- or post-pulses and background noise less than 90%, which proves the reliability of the result. Therefore, the temporal contrast decreases by approximately five times when using a single stage of the anti-saturated absorption effect. The results of other pre- or post-pulses and background noise also confirm the five-fold TCR effect. We also tested a solid film material that had good optical limiting properties,[57] but light scattering was a problem. Therefore, solid materials with a high damage threshold and good optical limiting properties should be found which will improve the performance of the TCR method.

*3.2.2. Optical Kerr effect*

In addition to using an anti-saturated absorption material, the OKE had also been used in optical limiting.[49] This implies that we can also utilize the OKE to decrease temporal contrast.

For any isotropically transmitted Kerr material, the refractive index of the material can be denoted as $n = n_0 + n_2 I$, where $n_0$ and $n_2$ are the linear and nonlinear refractive indices, respectively, and *I* is the intensity of the input pulse. As for a laser beam with a Gaussian spatial profile, the refractive index can be described as $n(r) =$

$n_0 + n_2 I(r) = n_0 + n_2 e^{-gr^2}$, where *r* is the radius of the beam, and *g* is the Gaussian index. This equation indicates that a laser-induced Kerr lens will be formed if the Gaussian beam is focused in an isotropically transmitted Kerr medium, which will result in self-focusing with $n_2 > 0$ for most media.

In general, the main peak pulse is several orders of magnitude stronger than its sub-pulses and background noise in the temporal domain. This means that the pre- or post-pulse and background noise will not induce the Kerr lens when the main peak pulse induces a strong nonlinear Kerr lens to the laser beam. Furthermore, for a Gaussian beam, OKE induced self-focusing is an ultrafast response process within hundreds of femtoseconds,[58] which means that the main peak pulse may induce a transient lens for itself, while the pre- or post-pulses and background noise will not be affected by the induced Kerr lens. The main pulse and the rest of the pre- or post-pulses will then be separated in the direction of propagation or in the spatial domain after a suitable distance. The main peak pulse on the external part of the beam will self-focus onto the central part of the beam, while the pre- or post-pulses and background noise will still be located on the external part of the beam. In this way, the temporal contrast of the laser beam for the external part of the beam will decrease after self-focusing occurs in the optical Kerr medium.

As shown in **figure 6**, approximately 4.0 mJ of a femtosecond laser pulse at 800 nm, output from a Ti:sapphire regenerative CPA amplifier (Legend Elite Series, Coherent) with a nice Gaussian spatial profile, was guided into the device. The reflected beam after BS with a reflective-to-transmissive ratio of approximately 7:93 was used for reference pulse generation through the SD process as before. After passing through a 0.5 mm thick fused silica plate P, a transmitted beam of approximately 3.8 mJ was focused by a cylindrical lens with a focal length of 1000 mm into a 1 mm thick fused silica plate OK, which was located approximately 30 mm behind the focal point of the cylindrical lens.

The beam transverse profiles were recorded in the same position after focusing on the beam path using a CCD (BC106, Thorlabs) with (**figure 7** (b)) or without (figure 7 (a)) a 1 mm thick fused silica glass plate. It should be noted that the CCD was operated in an auto exposure time mode. The self-focusing effect in the 1 mm thick fused silica plate lead to a 6.2 mm to 3.6 mm change in the FWHM size in the horizontal direction of the output beam. This is because the main peak pulse was self-focused to the center of the beam. Then the laser beam on one edge of the beam in the horizontal direction (figure 7 (b)) after the self-focusing effect was filtered by a home-made optical block B, shown in figure 6. It can then be concluded that the filtered beam has a TCR temporal contrast.

In this experiment, the temporal contrast of the test pulse without the 1 mm thick glass plate was also measured using both the commercial third-order cross-correlator (Sequoia 800) and our device based on the SRSI-ETE method shown in figure 6. The results are shown in figure 7 (c), where the black solid line and the blue solid line are measured using Sequoia 800 and the home-made SRSI-ETE, respectively. A good correlation is achieved for both measurements, which once again confirms the reliability of our home-made SRSI-ETE device.

The filtered beam (the area marked by the white dash square in figure 7 (a) and (b)) after self-focusing was then characterized by the home-made SRSI-ETE device shown in figure 6. As before, the calibrating pulse was introduced at approximately $t = 5$ ps using a 0.5 mm thick fused silica plate P. After analyzing the tilted spectral interferogram data obtained by the 2D imaging spectrometer, the temporal contrast of the TCR test pulse is achieved, shown in figure 7 (c) with a red solid line. This correlates with the change in the intensity of the calibrating pulse, which varied from $9.20 \times 10^{-4}$ to $1.56 \times 10^{-2}$ with the absence or presence, respectively, of the 1 mm thick fused silica plate in the optical path. In both measurements, the intensity varied approximately 17 times. The pre- or post-pulses and background noise from -25 ps to 10 ps for both characterizations also confirm the 17-fold TCR. Therefore, the TCR-based SRSI-ETE method using OKE can improve the dynamic range of the temporal contrast measurement by 17 times to approximately $10^9$ using one stage of the OKE process. This method can also be extended to other single-shot temporal contrast measurement methods, such as a cross-correlator, to improve its dynamic-range of characterization.

## 4. Conclusion

Single-shot characterization of the temporal contrast of a PW laser pulse with a high dynamic-range is important for understanding where the pre- or post-pulses and background noise are coming from in a PW laser facility, thus improving its temporal contrast. During ultra-high intensity laser physics experiments, a clear knowledge of the temporal contrast of the driving laser pulse will help to accurately explain the experimental results. A new idea has been proposed to improve the dynamic-range of characterization of a single-shot temporal contrast measurement using a novel TCR method. As proof-of-principle experiments, pulse stretching, anti-saturated absorption, and OKE were used to decrease the temporal contrast of the test pulse. Pulse stretching was found to be very simple with approximately two orders of magnitude TCR capability for laser pulses of tens of femtoseconds. With a single stage of the anti-saturation absorption process using the LBDBP sample, the temporal contrast was reduced approximately five-fold. Therefore, new anti-saturation absorption material with a higher TCR capability is expected to be found in the future. In addition, self-focusing based on OKE also showed great potential for the proposed single-shot TCR temporal contrast measurement method. Combined with a home-made SRSI-ETE device, this method increased the dynamic-range by approximately 17 times to approximately $10^9$ with a 1 mm thick fused silica glass plate. In combination with the single-shot cross-correlator techniques, one can achieve a dynamic-range up to $10^{11-12}$ using the proposed TCR method. Furthermore, it is expected that an even higher dynamic-range can be characterized by the combination of several techniques and cascaded TCR processes.

**Acknowledgement:**

The authors would like to thank Prof. Yuxia Zhao from Technical Institute of Physics and Chemistry, Chinese Academy of Sciences for supporting the LBDBP sample. This work is supported by the National Natural Science Foundation of China (NSFC) (grants 61527821 and 61521093), the Instrument Developing Project of the Chinese Academy of Sciences (grant YZ201538), the Strategic Priority Research Program of the Chinese Academy of Sciences (grant XDB16), and Shanghai Municipal Science and Technology Major Project (grant 2017SHZDZX02). Xiong Shen and Peng Wang contributed equally to this work.

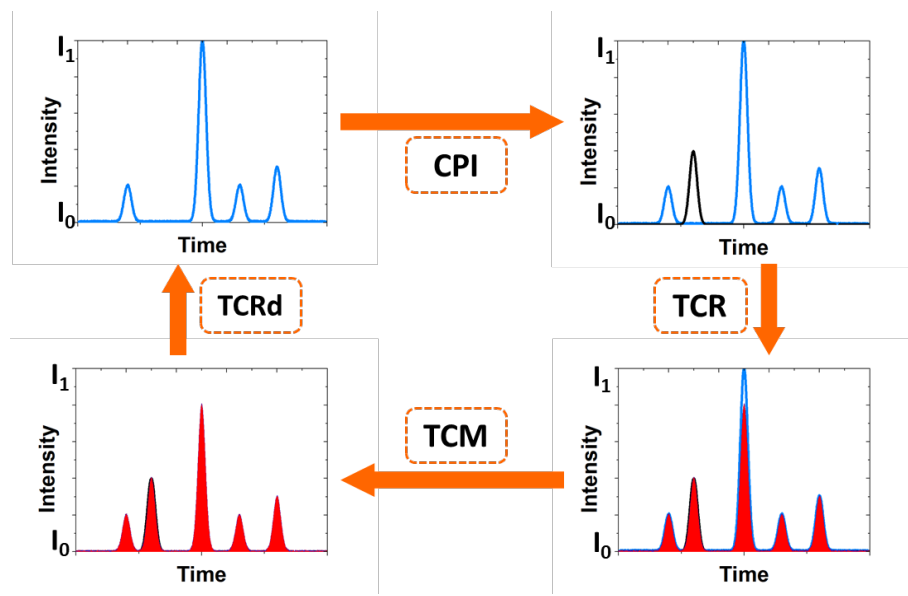

**Figure 1.** (Principle of high dynamic temporal contrast characterization. CPI, calibrating pulse introduction; TCR, temporal contrast reduction; TCM, temporal contrast measurement; TCRd, temporal contrast reconstruction)

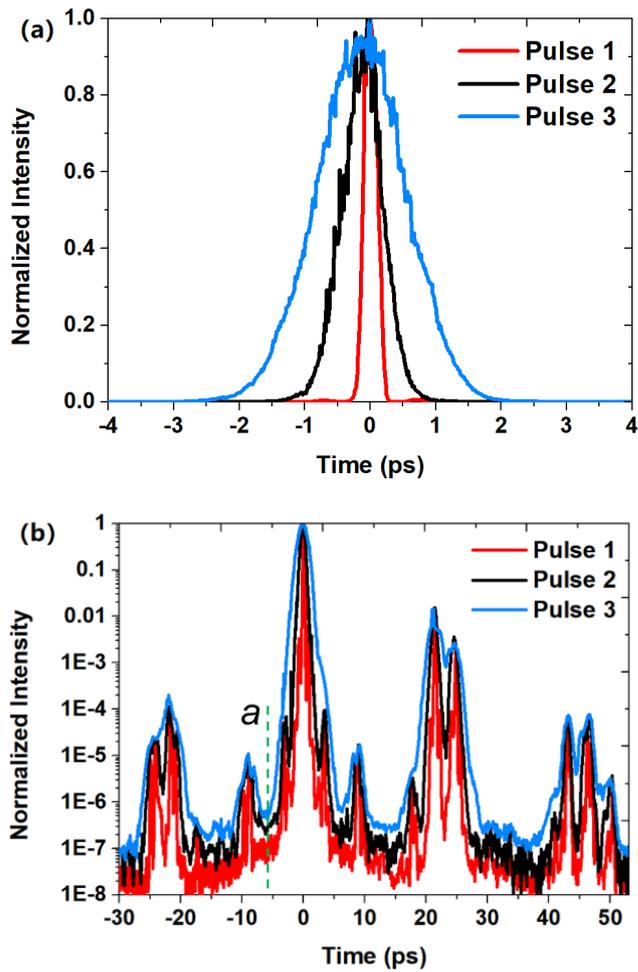

**Figure 2.** ((a) Third-order autocorrelation curves and (b) temporal contrast curves of all three pulses at three different pulse widths.)

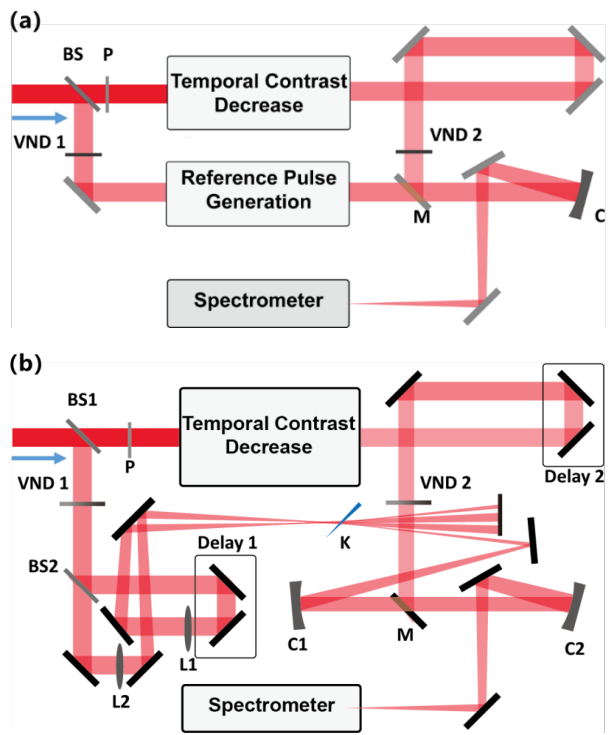

**Figure 3.** ((a) Diagram of a TCR-based SRSI-ETE device. (b) Diagram of a TCR-based SRSI-ETE device with reference pulse generation using the SD process.)

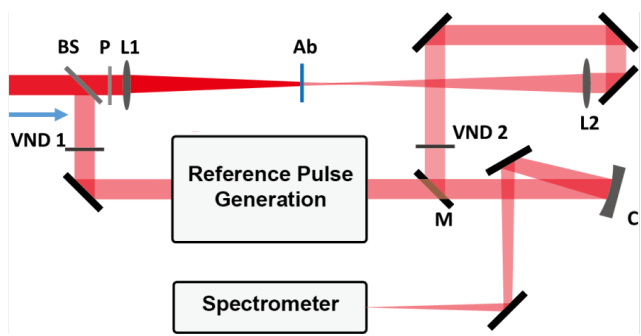

**Figure 4.** (Experimental setup for TCR using the optical limiting effect. P is a 0.5 mm thick fused silica plate. Ab is an anti-saturated absorption material. The scheme of the "Reference Pulse Generation" section is similar to the SD process shown in figure 3 (b).)

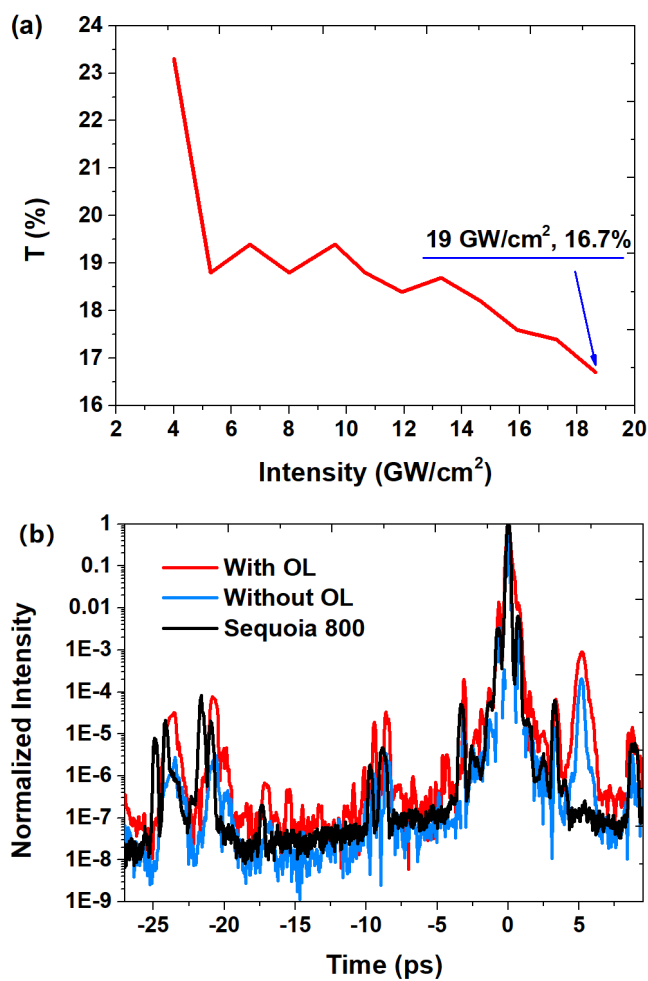

**Figure 5.** ((a) Anti-saturated absorption curve of LBDBP. (b) Temporal contrast signals of the input pulse, characterized by Sequoia 800 (black) or by our home-made SRSI-ETE device, shown in figure 4, without (blue) and with optical limiting effect (red) in the LBDBP sample.)

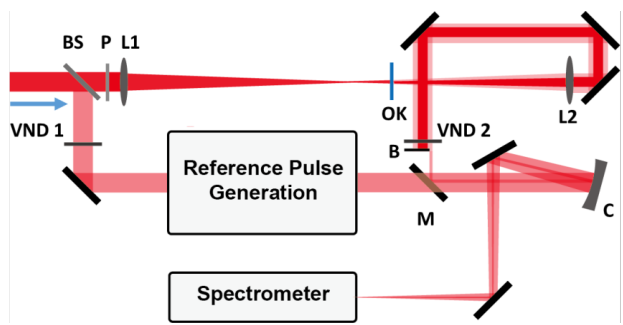

**Figure 6.** (Experimental setup for TCR using the optical Kerr effect. P, a 0.5 mm thick fused silica plate for generating reference calibrating pulses. OK, a 1 mm thick fused silica plate, used for self-focusing. The scheme of the "Reference Pulse Generation" section shows the same SD process as shown in figure 3 (b).)

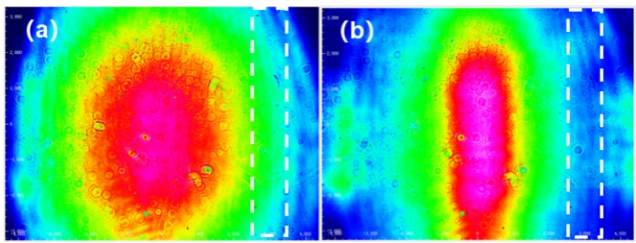

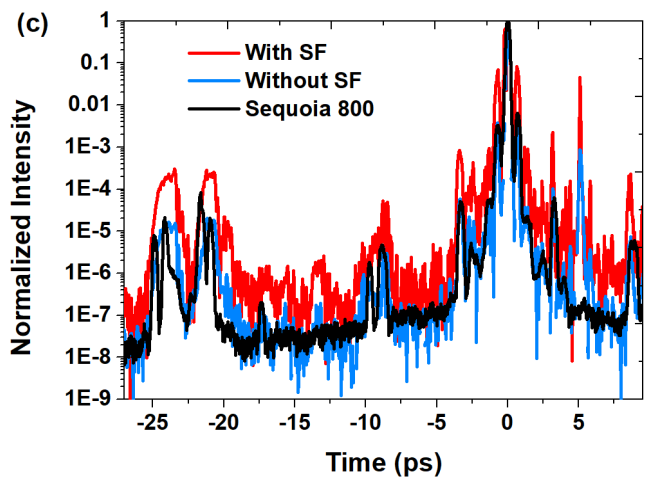

**Figure 7.** (Transverse profiles of beams without (a) or with (b) optical Kerr effect. (c) Temporal contrast curves of the input pulse, characterized by Sequoia 800 (black) or our device without (blue) and with the OKE (red) in a 1 mm fused silica plate.)